\documentclass{aa}

\usepackage{color}
\usepackage{comment}
\usepackage{natbib}
\usepackage{url}
\usepackage{hyperref}
\usepackage{subfigure}
\usepackage{xspace}
\usepackage[percent]{overpic}
\usepackage{subfloat}
\usepackage{subcaption}
\usepackage{graphicx}
\usepackage{caption}
\usepackage{subcaption}
\usepackage{afterpage}
\usepackage{float}
\usepackage{multicol} 
\usepackage{adjustbox} 

\usepackage{txfonts}
\newcommand{\src}{EXO~2030+375}
\newcommand{\hxmt}{{\it Insight}-HXMT\xspace}
\newcommand{\ixpe}{{IXPE}\xspace}

\begin{document}

   \title{Timing and spectral studies of the Be/X-ray binary EXO~2030+375 using \hxmt observations}

   \author{Yu-Jia~Du \inst{\ref{in:Tub}}
          \and Lorenzo~Ducci \inst{\ref{in:Tub},\ref{in:Genf}}
          \and Long~Ji \inst{\ref{in:SYSU}}
          \and Qing-Cui~Bu \inst{\ref{in:CCNU}}
          \and Ling-Da~Kong \inst{\ref{in:Tub}}
          \and Peng-Ju~Wang \inst{\ref{in:Tub}}
          \and Youli~Tuo \inst{\ref{in:Tub}}
          \and Andrea~Santangelo \inst{\ref{in:Tub}}
          }

   \institute{Institut f\"ur Astronomie und Astrophysik, Universit\"at T\"ubingen, Sand 1, D-72076 T\"ubingen, Germany \label{in:Tub}\\
   \email{du@astro.uni-tuebingen.de}
   \and
   ISDC Data Center for Astrophysics, Université de Genève, 16 chemin d’Écogia, 1290, Versoix, Switzerland \label{in:Genf}
   \and
       School of Physics and Astronomy, Sun Yat-Sen University, Zhuhai, 519082, People's Republic of China \label{in:SYSU}
    \and
       {Institute of Astrophysics, Central China Normal University, Wuhan 430079, P.R. China \label{in:CCNU}}
}

  \abstract{
   We report the X-ray spectral and timing analysis of the high mass X-ray binary EXO~2030+375 during the 2021 type-II outburst based on the \hxmt observations. Pulsations can be detected in the energy band of 1--150\,keV. The pulse profile shows energy and luminosity dependence and variability. We observed transitions in the pulse profile shape during the rising and the decaying phase of the outburst. The pulse fraction exhibits an anti-correlation with luminosity and a non-monotonic energy dependence, with a possible dip near 30\,keV during the outburst peak. The hardness-intensity diagrams (7--10\,keV/4--7\,keV) suggests state transitions during the early and late phases of the outburst. These transitions are consistent with the luminosity at which the pulse profile shape changes occur, revealing the source reaching the critical luminosity and transitioning between super-critical and sub-critical accretion regimes. We performed the average and phase-resolved spectral analysis, where the flux-resolved average spectra show a stable spectral evolution with luminosity. The phase-resolved spectral analysis reveals that the dependence of spectral parameters on the pulse phase varies with different luminosities.}


   \maketitle
%

\section{Introduction} \label{section_1}
Be/X-ray binary (BeXRB) systems consist of an accreting neutron star (NS) characterized by B $\sim$ $10^{12}\,$G and a O/B spectral main sequence or giant donor star {(\citealt{1976Natur.259..292M,2011Ap&SS.332....1R})}. In these systems, the NS can accrete matter supplied by the companion star primarily {via the stellar wind or the circumstellar disc}, thereby producing emission in the X-ray domain. The apparent luminosity covers many orders of magnitude from $\sim$ $10^{32}\,$erg\,$\rm{s}^{-1}$ up to $\sim$ $10^{41}\,$erg\,$\rm{s}^{-1}$ (see a recent review by \citealt{2023hxga.book..138M}). Besides the periodic (type-I) X-ray outbursts characterized by lower luminosity of {roughly} $L_{\rm x}$ < $10^{37}\,$erg\,$\rm{s}^{-1}$ often observed at the periastron passage of the NS, giant outbursts (type-II) are also observed in BeXRBs with peak luminosity of $L_{\rm x}$ $\ge$ $10^{37}$\,erg\,$\rm{s}^{-1}$ \citep{2011Ap&SS.332....1R}.

{In these binary systems}, the plasma is channeled along the magnetic field {of the neutron star} and accretes onto the polar cap, where the gravitational energy of the flow is released. 
It is believed that at luminosities greater than the critical luminosity ($L_{\rm crit}\approx 10^{36-37}$~erg~s$^{-1}$), the plasma flow is {decelerated} above the NS surface by the radiative pressure, {leading to the formation of a radiatively dominated shock above the polar cap and the creation of an extended emission region, known as the accretion column.} \citep{1976MNRAS.175..395B, 2012A&A...544A.123B}. The pulse profiles and X-ray spectra of accreting pulsars show different properties below and above $L_{\rm crit}$.
For example, the shapes of the pulse profile of many pulsars show a significant variability around $L_{\rm crit}$ (\citealt{2018MNRAS.479L.134T, 2020MNRAS.491.1851J, 2022ApJ...935..125W}).
Another observational tool typically used to infer properties about the different accretion regimes around $L_{\rm crit}$
is provided by cyclotron resonant scattering features (CRSFs), which are absorption-like lines in the continuum spectrum, produced by the quantized motion of electrons in strong magnetic fields.
CRSFs provide a direct measurement of the magnetic field, $E_{\text{cyc}} \approx 11.6 \times B_{12} \times (1 + z)^{-1} \, \text{keV} $ \citep{2019A&A...622A..61S}, where $B_{12}$ is the surface magnetic field strength in units of $10^{12}$\,G, $z$ is the gravitational redshift due to the NS mass and radius.
The energy of the fundamental CRSF shows a correlation with luminosity below $L_{\rm crit}$ and an anti-correlation above it (\citealt{2023hxga.book..138M}, and references therein).
Studies on the transition occurring around $L_{\rm crit}$ between different accretion regimes are crucial to improving our understanding of the dynamic of accretion flow geometry and radiation mechanisms in pulsars with high magnetic fields.
In this work, we present a spectral and timing analysis of one of these pulsars, EXO~2030+375, using 1$-$150~keV data from the \hxmt\ satellite obtained during the 2021 giant outburst.
\src\ is a transient BeXRB discovered by the {\it EXOSAT} observatory during a type-II outburst in 1985. It hosts a $\sim 42$\,s pulsar orbiting around the companion star in a rather eccentric orbit ($P_{\rm orb} \approx 46$~d, $e\sim0.41$; \citealt{1985IAUC.4066....1P, 1989ApJ...338..359P}. Its distance has been estimated as $2.4_{-0.4}^{+0.5}$\,kpc by the {\it Gaia} Data Release 3 \citep{2021AJ....161..147B}.

\begin{figure}[h!]
\centering 
\includegraphics[width=0.98\columnwidth]{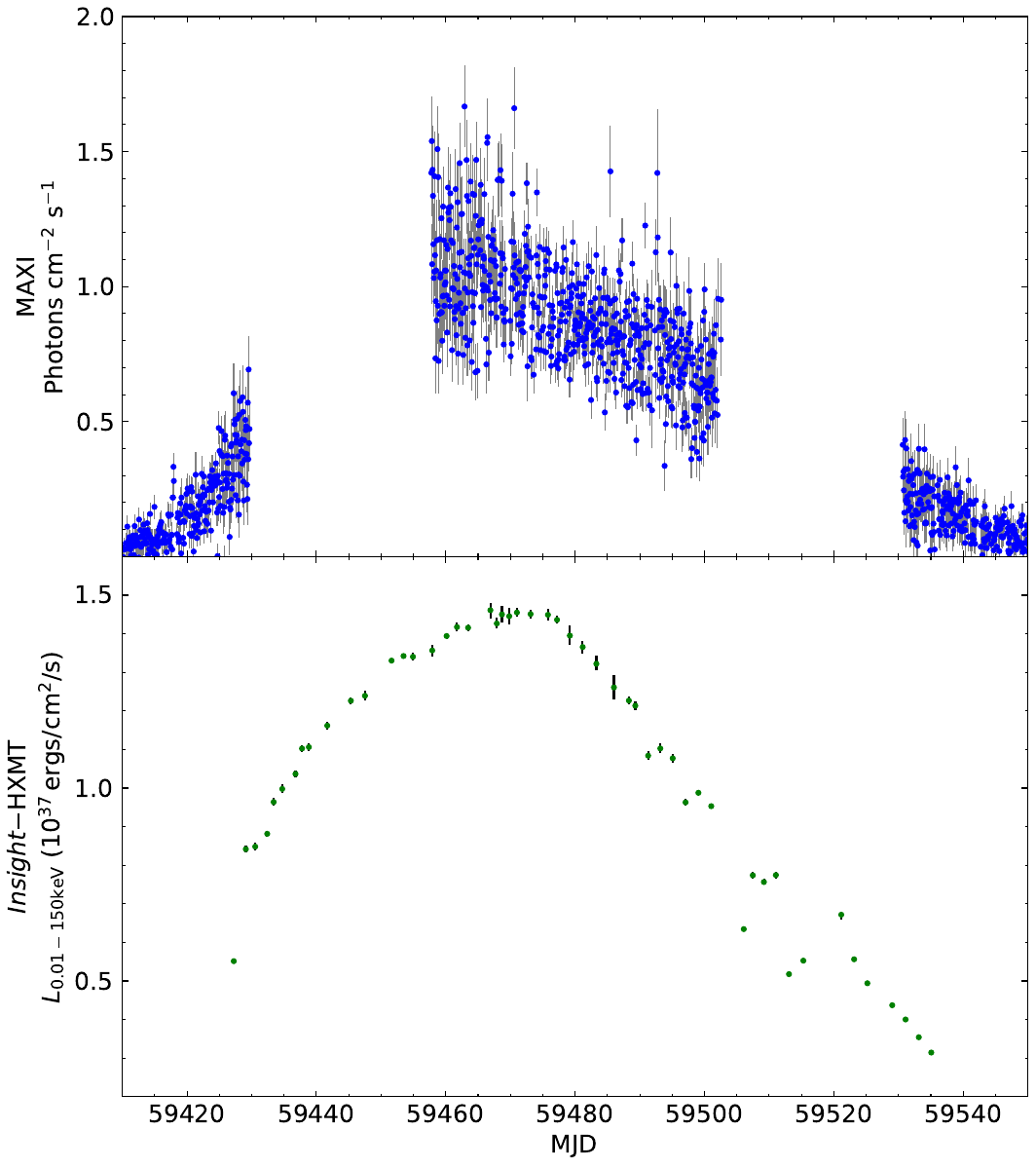}
\caption{ Light curve of \src\ as measured by MAXI (2$-$20~keV), {with data binned over one orbital revolution of the International Space Station, where MAXI is mounted ($\sim 1.5$~hours), shown in the first panel}. The second panel shows the light curve from the \hxmt\ observations during the 2021 giant outburst. Each point represents the daily-averaged luminosity (0.01$-$150\,keV), derived through spectral fitting of the \hxmt\ data. 
\label{fig:History}}
\end{figure}

Various interesting features have been observed in the broad band spectrum of \src. During a type-I outburst in 1996 observed by {\it RXTE}, \citet{1999MNRAS.302..700R} reported the possible detection of CRSF with a centroid energy of $\sim$ 36~keV. \src\ was observed to undergo a second type-II outburst in 2006. A broad Gaussian emission feature around 15~keV has been observed \citep{2007A&A...464L..45K}. A possible CRSF was reported at $\sim$ 63~keV in phase-resolved spectra extracted during the peak of the 2006 outburst \citep{2008A&A...491..833K}. \emph{Suzaku} observations of \src\ during type-I outbursts in 2007 and 2012 did not reveal the presence of CRSFs in the X-ray spectrum \citep{2016xnnd.confE..44F}. The third giant outburst was detected in 2021. \citet{2023MNRAS.521..893F} reported the evolution of the pulse profile with luminosity and provided an updated set of orbital and temporal parameters of \src\ with \hxmt. 
During a 2022 type-I outburst, \citet{2023A&A...675A..29M} used \textit{IXPE} to find that the polarization degree ranged from $0\%$ to $3\%$ in phase-averaged analysis and from $2\%$ to $7\%$ in phase-resolved analysis. Their combined polarimetric, spectral, and timing analyses revealed a complex accretion geometry for EXO 2030+375, significantly influenced by asymmetric magnetic multipoles and gravitational light bending.

In this work, we present the spectral and timing analysis of \src, using the data of \hxmt during the 2021 giant outburst. The observations and data processing are given in Sect. \ref{section_2}. We present the dependence of the pulse profile with time and energy. We analyzed phase-averaged, phase-resolved spectra of the 2021 outburst to explore spectral variations. The results are given in Sect. \ref{section_3}. In Sects. \ref{section_4} and \ref{section_5}, we discuss and summarize our results. Additionally, we provide supplementary results that support our main findings in the appendix.

\section{Observations and data reduction} \label{section_2}

The Hard X-ray Modulation Telescope (Insight-HXMT, also known as HXMT; \citealt{2020SCPMA..6349502Z}), launched on 15th June 2017, is the first Chinese X-ray astronomical satellite.
HXMT excels in its broad energy band (1$-$250~keV) and a large effective area in the hard X-rays energy band \citep{2020SCPMA..6349502Z}. \src\ was observed by \hxmt from 2021 July 28 (MJD 59423) to November 21 (MJD 59539). {Figure \ref{fig:History} presents the light curves of \src\ during the 2021 giant outburst. The top panel displays the 2–20 keV light curve observed by MAXI \footnote{The MAXI data were retrieved from \url{http://maxi.riken.jp/star_data/J2032+376/J2032+376.html}.} \citep{2009PASJ...61..999M}, while the bottom panel shows the light curve derived from the \hxmt\ observations through spectral fitting. }  There are 65 observations of the proposal P0304030 and 15 observations of the proposal P0404147, with a total exposure time of $\sim$ 2460\,ks. Especially, observations of the proposal P0404147 were at the peak of the outburst. {The observations utilized for the timing and spectral analysis of the source can be found in Appendix \ref{sec:observations} (Table \ref{tab:HXMTobsID_rise&decay})} The exposures were combined by date to enhance the statistics using \texttt{addspec} in {\tt heasoft} (version 6.31.1). 
The detectors of three payloads (LE: 1$-$15~keV, 384 cm$^{2}$; ME: 5$-$30~keV, 952 cm$^{2}$; HE: 20$-$250~keV, 2100 cm$^{2}$) on board \hxmt were used to generate the events in good time intervals (GTIs). The time resolution of the HE, ME, and LE instruments are $\sim$25~$\mu$s, $\sim$280~$\mu$s, and $\sim$1 ms, respectively. Data from \hxmt were considered in the range 2$-$150~keV, with the exclusion of 21$-$23~keV data due to the presence of an Ag feature \citep{2020JHEAp..27...64L}. Insight-HXMT Data Analysis software\footnote{\url{http://hxmtweb.ihep.ac.cn/}} (HXMTDAS) v2.04 and HXMTCALDB v2.04 were used to analyze the data. We screened events for three payloads in HXMTDAS using \texttt{legtigen}, \texttt{megtigen}, \texttt{hegtigen} tasks according to the following criteria for the selection of GTIs: (1) pointing offset angle $<$ 0.1\degr; (2) the elevation angle $>$10\degr; (3) the geomagnetic cut-off rigidity $>$8 GeV; (4) the time before and after the South Atlantic Anomaly passage $>$300 s; and 
(5) for LE observations, pointing direction above bright Earth $>$30\degr. We selected events from the small field of views (FoVs) for LE and ME observations, as well as from both small and large FoVs for HE observations due to the limitation of the background calibration. HXMT is a non-imaging telescope, the background is generated using a background model constructed through training and calibration on multiple calibration sources of the blind detectors, which are covered by tantalum lid. 
The software for background realization is incorporated in  \texttt{lebkgmap} \citep{2020SCPMA..6349505C}, \texttt{mebkgmap} \citep{2020SCPMA..6349504C}, and \texttt{hebkgmap} \citep{2020SCPMA..6349503L} implemented in HXMTDAS, respectively. The arrival times of photons at HXMT are corrected to the Solar system barycentre by \texttt{hxbary} in HXMTDAS. The correction of the arrival times to account for the orbital modulation was performed using the orbital ephemeris reported by \citet{2023MNRAS.521..893F}. The XSPEC v12.13.0c software package \citep{1996ASPC..101...17A} was used to perform the spectral fitting. We included renormalisation constant factors in all the spectral ﬁttings to account for inter-calibration uncertainties between the three \hxmt\ instruments. The uncertainty estimated for each spectral parameter is for 90\% confidence, and a systematic error of 1\% is added to LE , ME, and HE spectrum. \\

\begin{figure}[htbp]
  \centering
  \includegraphics[width=0.94\columnwidth]{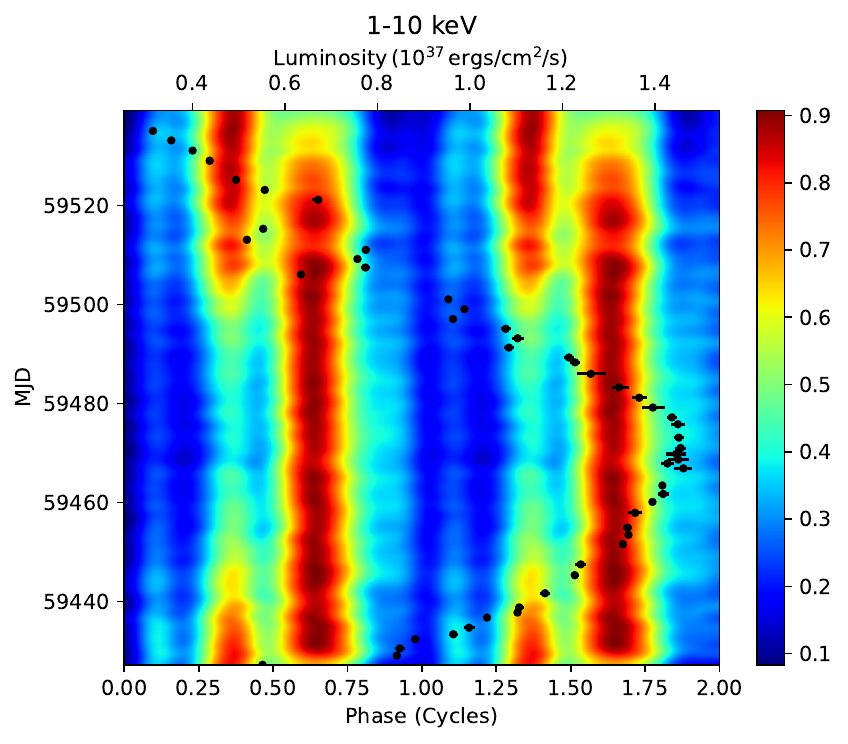}\\[1mm]
  \includegraphics[width=0.94\columnwidth]{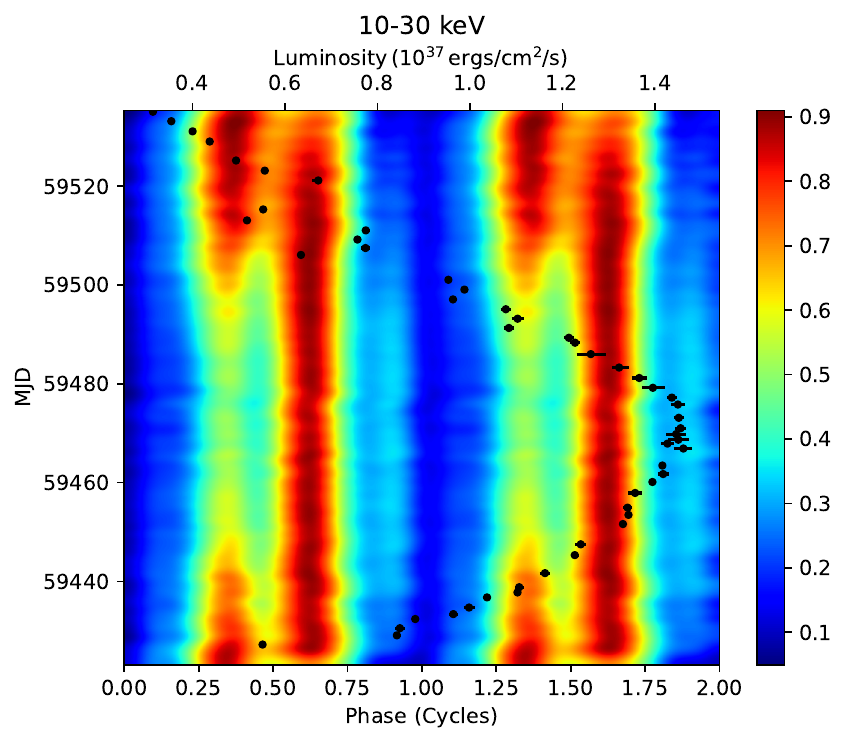}\\[1mm]
  \includegraphics[width=0.94\columnwidth]{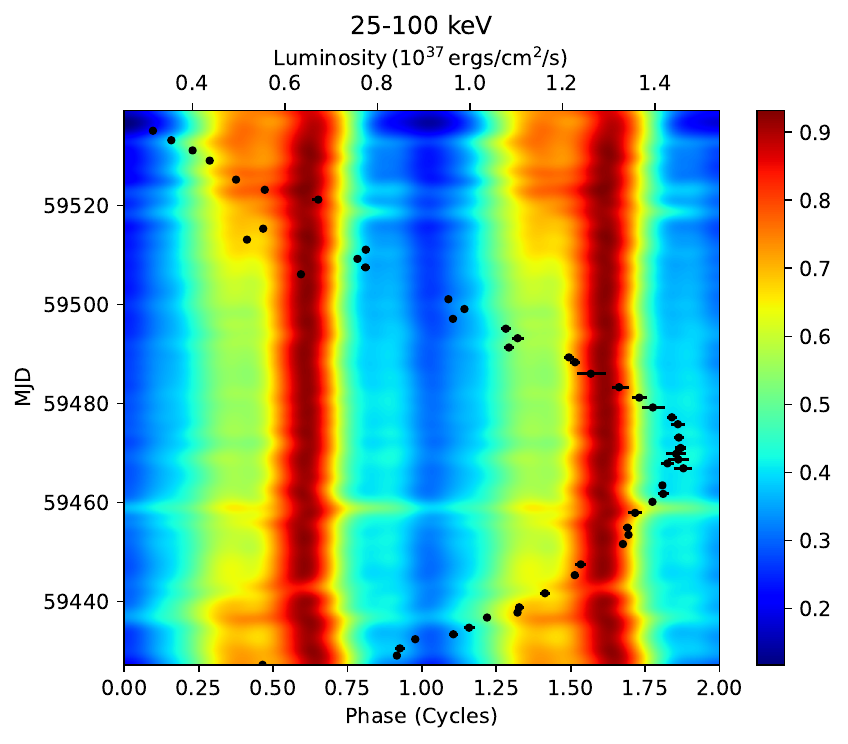}
  \caption{2D maps describe the evolution of the pulse profile with time observed by \hxmt during the 2021 outburst (proposal P0304030 and P0404147, MJD 59427-59535). Top: 1$-$10~keV; middle: 10$-$30~keV; and bottom: 25$-$100~keV. The black points in the figure represent the luminosity (0.01$-$150 keV).} 
  \label{p_vs_L} 
\end{figure}

\section{Results} \label{section_3}
\subsection{Timing analysis}

{The 0.01--150\,keV light curve of the 2021 outburst is presented in the second panel of Fig. \ref{fig:History}. Luminosities were derived from the  spectral analysis detailed in Sect.} \ref{section_average}.  The outburst reaches its maximum luminosity at MJD 59466. The duration of the rising phase of the outburst is approximately 40 days, spanning from MJD 59427 to MJD 59466, and covering a luminosity range of 0.55 $\times$ $10^{37}\,$erg\,$\rm{s}^{-1}$ to 1.45 $\times$ $10^{37}\,$erg\,$\rm{s}^{-1}$. The duration of the decay phase is approximately 69 days, spanning from MJD 59466 to MJD 59535, and covering a luminosity range of 1.45 $\times$ $10^{37}\,$erg\,$\rm{s}^{-1}$ to 0.31 $\times$ $10^{37}\,$erg\,$\rm{s}^{-1}$.

Figure \ref{p_vs_L} shows the temporal evolution of the pulse profiles of \src. {To enhance the statistics, data within one day was combined. To accurately measure the spin period for timing analysis, we performed epoch folding \citep{1987A&A...180..275L} on the extracted light curves. The profiles were obtained from light curves with a bin size of 0.5~s extracted in three energy bands (LE: 1--10~keV, ME: 10--30~keV, HE: 25--100~keV).} The pulse profiles were aligned, with phase zero defined as the minimum value of the pulse profile. 

A two-peak structure at lower flux can be clearly identified for the {1--10~keV and 10--30~keV
energy bands.} The main peak at $\sim$ 0.60  phase remains stable throughout the outburst, with its intensity decreasing near the end. Conversely, as the outburst rises, the intensity of the secondary peak at $\sim$ 0.35 phase decreases. When the luminosity exceeds $\sim 1.0 \times 10^{37}$erg $\rm{s}^{-1}$, the intensity of the secondary peak remains relatively low until the luminosity decreases back to $\sim 1.0 \times 10^{37}$erg $\rm{s}^{-1}$. At this point, the intensity of the secondary peak increases again, gradually becoming higher than that of the main peak, with the secondary peak eventually becoming dominant as the main peak's amplitude decreases. For HE, the secondary peak is not as prominent as it is for LE and ME; the main peak remains more dominant. However, higher intensity can still be observed at phase 0.3--0.5 at lower luminosities. In general, the secondary peak at $\sim$ 0.35 
 phase dominates at lower luminosities, while the main peak at $\sim$ 0.60 phase dominates at higher luminosities.

\begin{figure*}[!t]
  \centering
    \label{fig:a}
    \includegraphics[width=3in]{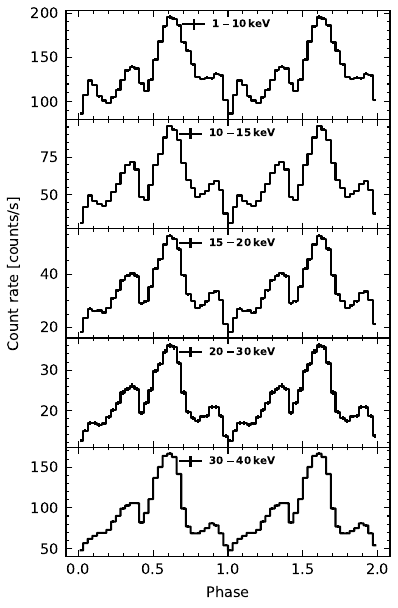}
  \hspace{0.1in}
    \label{fig:b}%
    \includegraphics[width=3in]{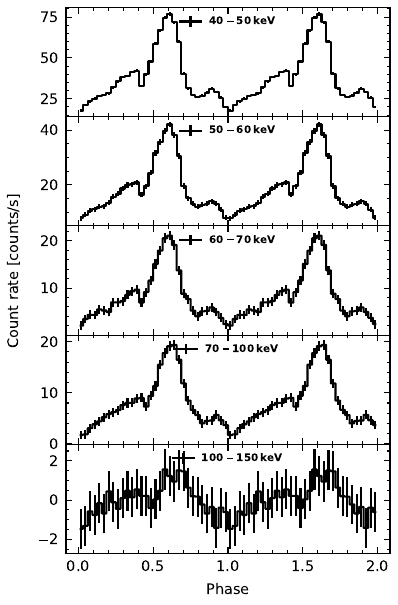}
  \caption{Energy-resolved pulse profiles of \src\ near the maximum of the outburst.}
  \label{PP_E} 
\end{figure*}

Figure \ref{PP_E} presents the energy-resolved (1$-$10, 10$-$15, 15$-$20, 20$-$30, 30$-$40, 40$-$50, 50$-$60, 60$-$70, 70$-$100, 25$-$100, and 100$-$150~keV) background-subtracted pulse profiles of the source obtained near the maximum of the 2021 outburst (ObsID P0404147, MJD 59460-59461). The four-peak pulse profile evolves to a single peak as the energy increases. The small peak at $\sim$ 0.1 phase gradually decreases with increasing energy and is no longer detectable above 30~keV. Another small peak at $\sim$ 0.9 phase gradually decreases when the energy exceeds 40~keV. The main peak at $\sim$ 0.60 phase was dominant at all energy bands while the secondary peak at $\sim$ 0.35 phase can only be detected below 70~keV. Pulsations can still be detected above 100~keV.

\begin{figure}
   \centering
      \includegraphics[width=8cm]{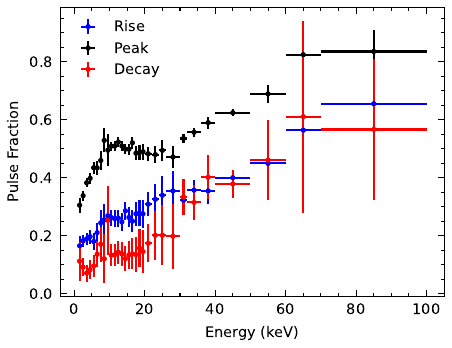}
      \caption{Variation in the PF with energy (keV). Rise: P0304030016, L $\sim$ 1.16 $\times$ $10^{37}\,$erg\,$\rm{s}^{-1}$. Peak: P0404147, and L $\sim$ 1.40 $\times$ $10^{37}\,$erg\,$\rm{s}^{-1}$. Decay: P0304030060, L $\sim$ 5.56 $\times$ $10^{36}\,$erg\,$\rm{s}^{-1}$.}
      \label{PF_E}
   \end{figure}

\begin{figure}
   \centering
      \includegraphics[width=8cm]{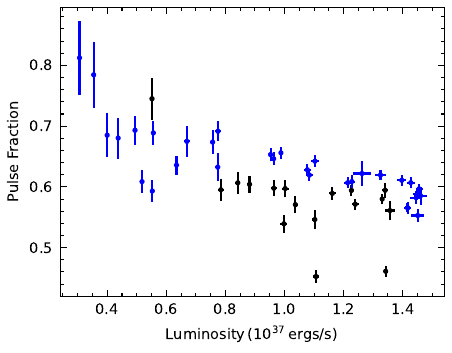}
      \caption{Variation in pulse fraction with luminosity in 25--100 keV energy range during the 2021 outburst.  Black crosses represent the rise of the outburst, blue the decay.}
         \label{PF_L}
   \end{figure}

We defined the pulse fraction (PF) parameter as the ratio of the difference to the sum of the maximum, $P_{max}$, and minimum, $P_{min}$, intensities of pulse profile, PF=$\frac{P_{max}-P_{min}  }{P_{max}+P_{min}}$. The variations in the PL with the energy at different luminosities during the rise (L $\sim$ 1.16 $\times$ $10^{37}\,$erg\,$\rm{s}^{-1}$), near the peak (L $\sim$ 1.40 $\times$ $10^{37}\,$erg\,$\rm{s}^{-1}$), and during the decay (L $\sim$ 5.56 $\times$ $10^{36}\,$erg\,$\rm{s}^{-1}$) are shown in Fig. \ref{PF_E}. The horizontal errors represent the energy ranges for which the PF is calculated, and the vertical errors indicate the corresponding error in PF measurements. {In general, the pulse fraction increases with energy, while a bump around 10\,keV was observed near the peak of the outburst.} Figure \ref{PF_L} shows the variation in the pulse fraction (25--100~keV) with the luminosity throughout the outburst, revealing an anti-correlation between the pulse fraction and luminosity during both the rise and decay phases.

\subsection{Phase-averaged analysis}\label{section_average}

   \begin{figure}
   \centering
      \includegraphics[width=8cm]{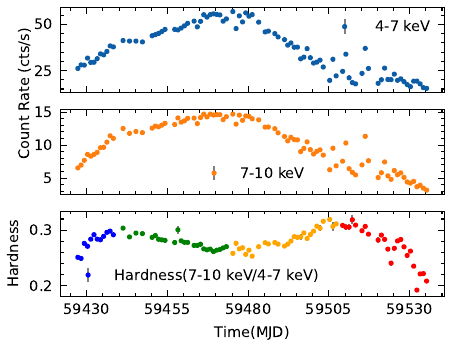}
      \caption{Net light curves and hardness of EXO 2030+375 of all observations from MJD 59427 to MID 59535. The hardness is defined as the ratio 7–10 keV / 4–7 keV. Each point corresponds to the data of one day. {The evolution of hardness over time is represented by dividing it into four distinct periods, each indicated by a different color: blue (MJD 59427 to 59438), green (MJD 59438 to 59473), orange (MJD 59473 to 59508), and red (MJD 59508 to 59535). }
}
         \label{HID_vs_t}
   \end{figure}

    \begin{figure}
   \centering
      \includegraphics[width=8cm]{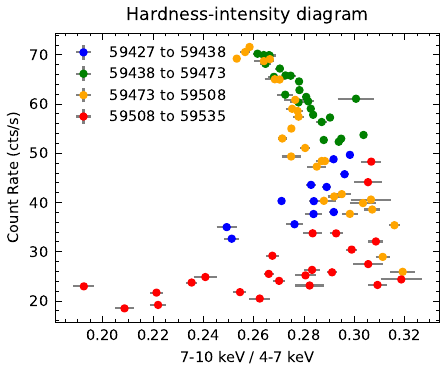}
      \caption{HID extracted from Insight- HXMT/LE data. Hardness is defined as the ratio 7–10 keV / 4–7 keV. {The colors are described in the same way as in Figure \ref{HID_vs_t}.}}
         \label{HID}
   \end{figure}

\begin{table}[!t]
\caption{Best-fit parameters of the phase-averaged broadband spectrum of \src\ as observed by \hxmt at the maximum of the outburst using the model \texttt{const$\times$tbabs(gauss+bbodyrad+gauss+cutoffpl)} in the 2--50~keV energy band.}
\label{table:peak_spectrum}

\centering
\begin{tabular}{p{5cm}p{2cm}}  
\hline\hline 
\noalign{\smallskip}
Parameter &  Value\\ 
\noalign{\smallskip}
 \hline 
\noalign{\smallskip}
$C_{\rm LE}$ (fixed) & 1 \\
$C_{\rm ME}$ & $0.929\pm0.006$ \\
$C_{\rm HE}$ & $0.882\pm0.011$ \\
$N_{\rm H}$\,[$10^{22}$\,cm$^{-2}$] & $2.32\pm0.13$ \\
$\Gamma$ & $0.942\pm0.025$ \\
Norm$_\Gamma$ & $0.565\pm0.002$ \\
$kT_{\rm bb}\,$[keV] & $0.74\pm0.03$ \\
Norm$_{bb}$ & $335.4\pm64.4$ \\
$E_{\rm cut}\,$[keV] & $18.3\pm0.5$ \\
$E_{\rm K\alpha}\,$[keV] & $6.69\pm0.02$ \\
$\sigma_{\rm K\alpha}\,$[keV] & $0.20\pm0.04$ \\
Norm$_{\rm K\alpha}$ & $0.005\pm0.001$ \\
$E_{\rm line}\,$[keV] & $5.71\pm0.09$ \\
$\sigma_{\rm line}\,$[keV] & $0.62\pm0.12$ \\
Norm$_{\rm line}$ & $0.012\pm0.002$ \\
Flux$_{\rm 0.01-150~keV}$ & $2.239\pm0.024$ \\
$\chi^2$/d.o.f. & $938.85/1250$ \\
\hline
\end{tabular}
\tablefoot{
{Normalization of the power law in units of photon\,cm$^{-2}$\,s$^{-1}$~keV$^{-1}$ at 1~keV.}
{Unabsorbed flux (in units of $10^{-8}\,$erg\,cm$^{-2}$\,s$^{-1}$) calculated for the entire model, obtained using the \texttt{cflux} command from \textsc{xspec} as resulting from the \textit{HXMT} data. Uncertainties are given for a 68\% confidence level.}
}
\end{table}

\begin{figure}[!t]
\centering
\includegraphics[width=9cm]{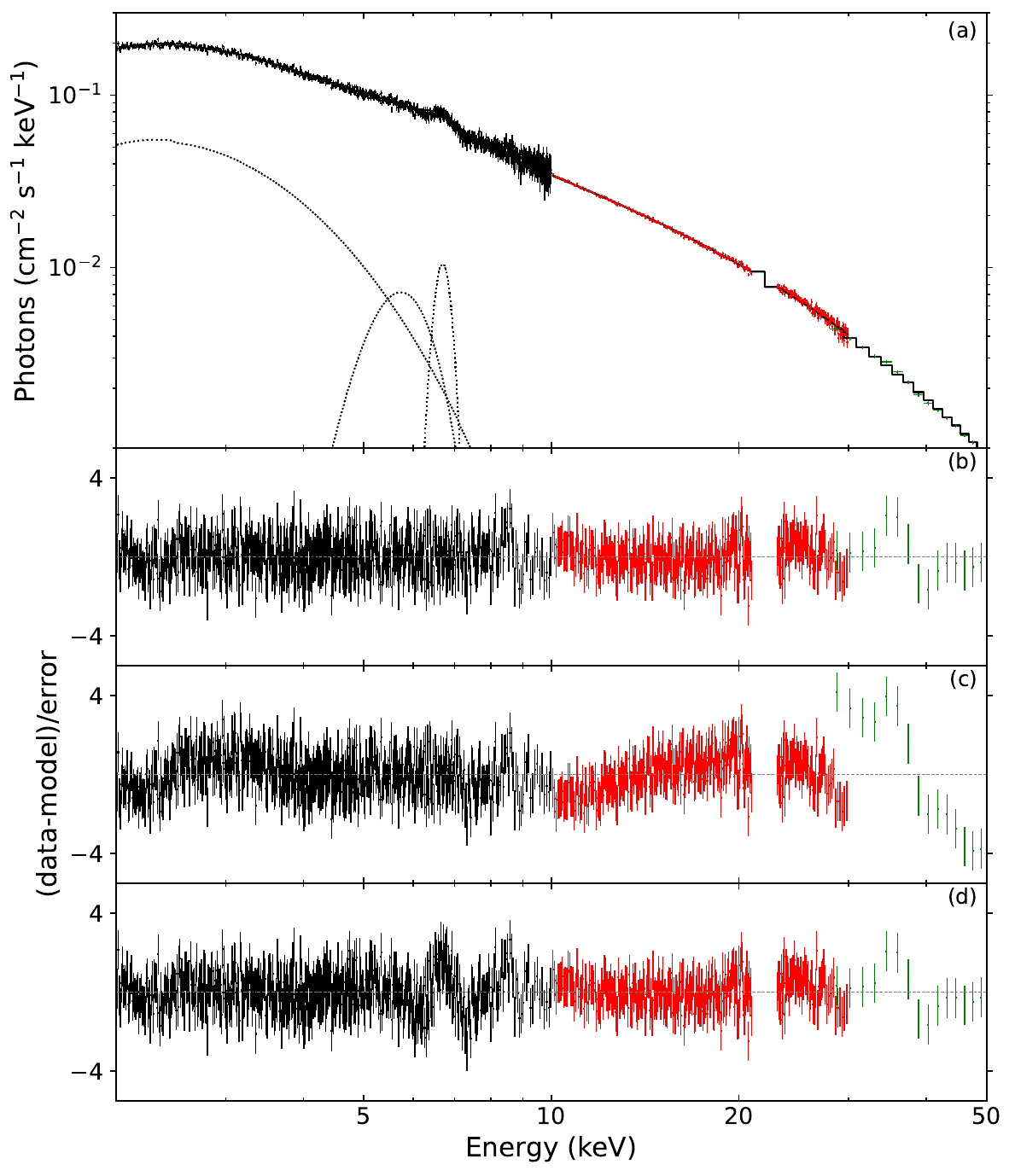}
\caption{Broadband spectrum of EXO 2030+375 at the peak of the 2021 giant outburst in energy range 2--50~keV using the model \texttt{const$\times$tbabs(gauss+bbodyrad+gauss+cutoffpl)}. (a) unfolded spectrum; {(b) residuals for the best-fit model; (c) residuals after fitting without the blackbody component; and (d) residuals after fitting without the feature around 5.7~keV.}
}
\label{average_spectrum}
\end{figure}

{The outbursts of accreting pulsars typically go through different spectral states, which can be diagnosed using a hardness-intensity diagram (HID) \citep{2013A&A...551A...1R}. HID shows how a source spectral hardness -- defined as the ratio of count per second (or flux) between two distinct X-ray energy bands -- varies with its total intensity.} Figure \ref{HID_vs_t} displays the net light curves in the energy ranges of 4--7~keV and 7--10~keV for all observations, as well as the hardness ratio of 7--10~keV / 4--7~keV versus time. The light curves are rebinned to one--day intervals. The hardness ratio is divided into four parts, represented by different colors, to illustrate its evolution over time. Figure \ref{HID} displays the HID.

Table \ref{tab:HXMTobsID_rise&decay} lists all the observations. We first focused on the observations near the peak of the 2021 outburst (MJD 59460-59461). {To enhance the counting statistics of the energy spectra, we combined seven exposures within a single day (see Table \ref{tab:HXMTobsID_rise&decay}: obs. IDs: 107\tablefootmark{a}, 108\tablefootmark{a}, 109\tablefootmark{a}, 110\tablefootmark{a}, 111\tablefootmark{a}, 112\tablefootmark{a}, and 113\tablefootmark{a}) using the \texttt{addspec} and \texttt{addrmf} tasks.} We restricted the energy bands for spectral analysis to 2–10~keV, 10–30~keV, and 28–50 keV~for LE, ME, and HE, respectively, due to the relatively higher background above 50~keV.

{Our choice of continuum model is based on earlier analyses of \src\ (e.g., \citealt{2007A&A...464L..45K}). The model consists of an absorbed cutoff power-law and a blackbody component, the latter included to account for the soft excess observed {below $\sim 5$~keV.}}

{In {\tt xspec}, we employed a cross-normalization constant to fit all instruments jointly and adopted the \texttt{tbabs} absorption model, setting the abundances to \texttt{wilm} \citep{2000ApJ...542..914W}. We added a Gaussian function centered around $\sim6.6$~keV, which likely corresponds to higher ionization states of iron, observed in higher luminosity observations (\citealt{2007A&A...464L..45K, 2016xnnd.confE..44F}).}

{Additionally, we included a second Gaussian component at $\sim5.7$~keV. This feature arises from discrepancies in single-split event count rates across the three LE detector boxes (private communication with the HXMT calibration team). Importantly, this feature does not impact the rest of the analysis or the main results. Panel (d) in Fig. \ref{average_spectrum} shows the residuals when the $\sim 5.7$~keV component is excluded from the spectrum: a residual feature appears at $\sim 6.16$~keV. This artifact occurs because the Gaussian component initially included to model the iron line attempts to fit the $\sim 5.7$~keV feature as well.}

{The best-fit model, consisting of an absorbed cutoff power-law with a blackbody and two Gaussian components, resulted in a $\chi^2$ value of 938.85 for 1250 degrees of freedom. The blackbody temperature was determined to be $kT=0.74 \pm 0.03$~keV, with the radius of the emitting region estimated at $4.4{+0.4\atop-0.5}$~km, assuming a distance of 2.4~kpc. The best-fit photon index was $0.942 \pm 0.025$, and the cutoff energy was $18.3 \pm 0.5$~keV. The interstellar absorption ($N_{\rm H}$) was found to be ($2.32\pm0.13$) $\times$ $10^{22}\,$ atoms cm$^{-2}$. Table \ref{table:peak_spectrum} provides a comprehensive summary of the spectral parameters derived from the best-fit model.}

{When the spectrum was fitted without the blackbody component, the $\chi^2$/d.o.f. increased to 1217.04/1252, and the residuals (Fig. \ref{average_spectrum}, panel c) worsened significantly. In this case, the best-fit required the constant parameter for the HE data (green points in Fig. 8) to shift to a much lower value compared to that listed in Table \ref{table:peak_spectrum}. If this constant is frozen at its previous value, the residuals exhibit a more pronounced wavy structure. To assess the statistical significance of the blackbody component, we performed an F-test, which yielded a probability of $10^{-70}$, strongly supporting the inclusion of the blackbody component in the model.
Throughout the analysis, the \texttt{cflux} convolution model was used to compute all unabsorbed flux values.}

  \begin{figure}
   \centering
      \includegraphics[width=8cm]{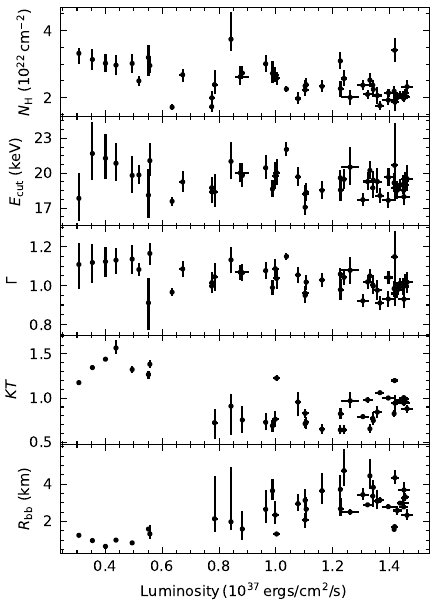}
      \caption{Flux-resolved parameters using all the observations of the 2021 giant outburst in energy range 2--50~keV. }
         \label{par_vs_L}
   \end{figure}

{The cutoff power-law with blackbody model discussed above was applied to fit the average spectrum for all the observations.} For ObsID P0304030051, P0304030052, P0304030054, P0304030055, and P0304030059, the inclusion of the blackbody component was unnecessary to achieve a good fit.
We also investigated spectral variability as a function of accretion luminosity. Given the smooth variation in the spectral parameters throughout the outburst, exploring correlations with X-ray flux is meaningful. The flux-resolved parameters are shown in Fig. \ref{par_vs_L}. The parameters involved are the equivalent hydrogen column density ($N_\mathrm{H}$), the photon index ($\Gamma$), the folding energy ($E_\mathrm{cut}$), the blackbody temperature ($kT_{\rm bb}\,$), and the size of the emitting region ($R_{\rm bb}$). The photon index ranges from 0.8 to 1.3, while the cutoff energy varies between 16~keV and 24~keV. Overall, the spectral continuum demonstrates stable evolution with luminosity. Initially, the blackbody temperature rises with luminosity until reaching $0.5 \times 10^{37}$~erg~s$^{-1}$, beyond which it decreases. The blackbody radius evolves inversely with the temperature.

\subsection{Phase-resolved analysis}

\begin{figure*}
  \centering
    \includegraphics[width=2.3in]{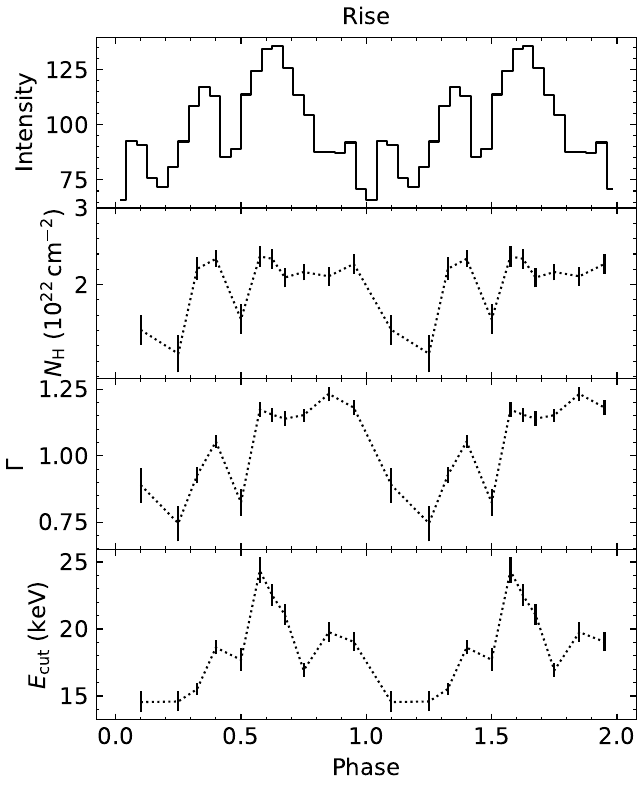}
    \label{fig:a}
  \hspace{0.1in}
    \includegraphics[width=2.3in]{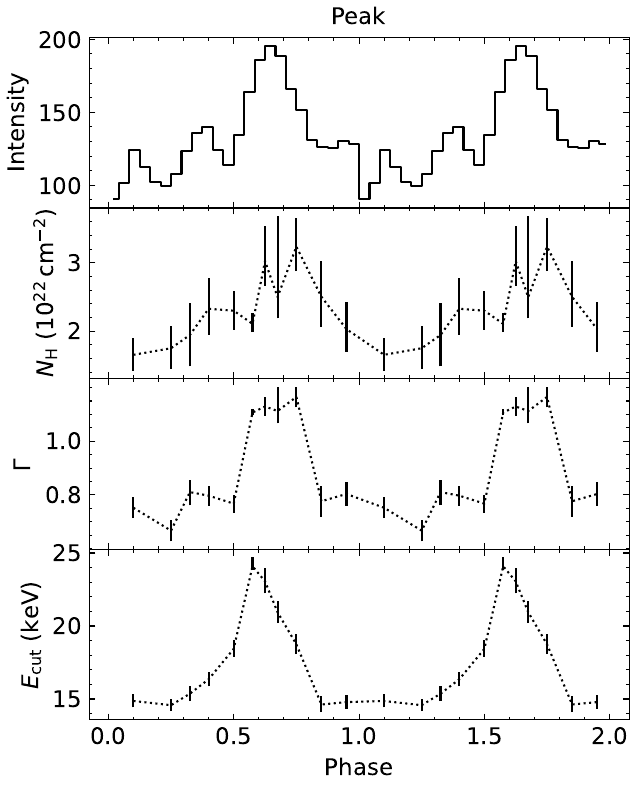}
    \label{fig:b}
  \hspace{0.1in}
    \label{(c)}
    \includegraphics[width=2.3in]{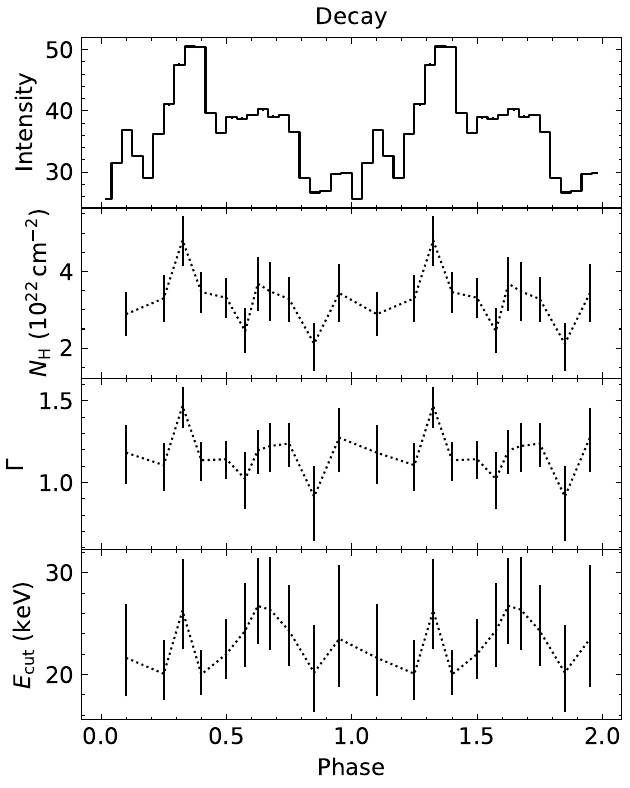}
    \label{fig:c}
  \caption{Modulation of phase-resolved parameters observed by \hxmt. Rise: MJD 59436-59438, L $\sim$ 1.0 $\times$ $10^{37}\,$erg\,$\rm{s}^{-1}$; peak: 59460-59461 MJD, L $\sim$ 1.45 $\times$ $10^{37}\,$erg\,$\rm{s}^{-1}$; decay: 59529-59533 MJD, L $\sim$ 0.4 $\times$ $10^{37}\,$erg\,$\rm{s}^{-1}$. {For each of the three subfigures, the panels (from top to bottom) show the pulse profile (1--10~keV), the column density ($N_{\rm H}$), the power-law photon index ($\Gamma$), and the cutoff energy ($E_{\rm cut}$).}}
  \label{prs} 
\end{figure*}

The phase-resolved spectral analysis of the outburst rise, peak, and decay were performed to explore variations of spectral parameters with pulse phase and different luminosities. Eleven phase bins were selected to sample different intensity levels shown by the pulse profile. {To detect the presence of a possible CRSF, we initially extracted spectra across a broader energy range, namely, from 2--100~keV. However, no definitive evidence of CRSF was observed. To minimize the impact of high background and maintain consistency in our spectral analysis, we ultimately narrowed the energy range to 2--50~keV when presenting our results (see Appendix \ref{sec:pulse-on-off} for more details).} To enhance the statistics, observations in the similar flux levels were combined: rise: 59436-59438 MJD, L $\sim$ 1.0 $\times$ $10^{37}\,$erg\,$\rm{s}^{-1}$;  peak: 59460-59461 MJD, L $\sim$ 1.40 $\times$ $10^{37}\,$erg\,$\rm{s}^{-1}$; decay: 59529-59533 MJD, L $\sim$ 0.4 $\times$ $10^{37}\,$erg\,$\rm{s}^{-1}$. The spectra obtained for eleven phase bins were fitted again with the same continuum model with the corresponding phase-averaged spectra. {During the rise and decay phases of the outburst, when the flux was relatively low, the spectral feature at approximately $\sim 5.7$~keV,  identified in the average spectrum near the peak of the outburst (Sect. \ref{section_average}), was not required to fit the data in individual phase bins. Consequently, the Gaussian component representing this feature was excluded from the best fit model, which for these parts of the outburst was: const×tbabs×(bbodyrad+gauss+cutoffpl).
At the peak of the outburst (59460$-$59461~MJD), the Gaussian component at $\sim 5.7$~keV was included to account for the observed feature, as in the average spectrum model. The resulting {\tt xspec} model for this interval was: const×tbabs×(gauss+bbodyrad+gauss+cutoffpl).} The best-fit phase-resolved continuum parameters as a function of the pulse phase are reported in Fig. \ref{prs}. To investigate the variation in the phase-resolved parameters with pulse profile morphology, we generated average pulse profiles for the LE (1--10~keV) energy band across the three time intervals. The LE band was chosen because the evolution of the main peak and the secondary peak with varying luminosities is most evident in this energy range. The disparity in the amplitude of the peaks is clearly observable.

For all the three time intervals, the power-law photon index shows a strong dependence on the pulse profile. Near the peak of the outburst, it reaches the highest at the main peak at phase $\sim$ 0.6. During the rise of the outburst, the photon index remains consistently high on the right flank of the main peak (phase 0.70-0.95). During the decay of the outburst, the amplitude of the main peak is lower and broader than that of the secondary peak at 0.35 phase. The photon index reaches its maximum value at the peak of the secondary peak. For the three time intervals, the cutoff energy varies with pulse phase, reaching its highest value at the peak of the main pulse.

\section{Discussion} \label{section_4}

We have studied spectral and timing characteristics of accreting pulsar \src\ in the 2021 outburst with \hxmt. During the observation, the X-ray luminosity ( 0.01--150~keV ) was between 0.31 $\times$ $10^{37}\,$erg\,$\rm{s}^{-1}$ and 1.45 $\times$ $10^{37}\,$erg\,$\rm{s}^{-1}$ assuming the distance of 2.4\,kpc. 

\citet{2013A&A...551A...1R} presented the hardness-intensity diagram (HID) of the 2006 outburst of \src\ (Fig. 3 in \citet{2013A&A...551A...1R}), where hardness is defined as the ratio of 7--10~keV to 4--7~keV. The diagram shows a significant transition at a luminosity of $\sim$ 4 $\times$ $10^{36}\,$erg\,$\rm{s}^{-1}$ (d=2.4\,kpc). The study indicates that for several X-ray pulsars, including \src, the evolution through the HID suggests state transitions during the early and late phases of the outburst. They demonstrate that these state transitions occur when a critical luminosity is reached. Additionally, the spectral continuum also shows a transition at this luminosity (Fig. 6 in \citet{2013A&A...551A...1R}). This indicates that in 2006, \src\ likely underwent two different accretion regimes. These regimes are probably associated with the mechanisms by which the accretion flow is decelerated in the accretion column: radiation pressure in the supercritical regime and Coulomb interactions in the subcritical regime \citep{2012A&A...544A.123B}. Using \hxmt data from the 2021 outburst, the HID diagram (7--10~keV/4--7~keV) shows transitions during both the rise and decay phases, indicating changes in the spectral composition. The transitions occur around MJD 59438 (rise phase) and MJD 59507 (decay phase) at $L_{\rm x}$ $\sim$ $10^{37}\,$erg\,$\rm{s}^{-1}$.

In the present study, the pulse profile has been observed to evolve with luminosity (Fig. \ref{p_vs_L}). The double-peak structure remains present throughout, with the intensity of the main peak showing minimal variation with luminosity, while the intensity of the secondary peak exhibits significant changes. As shown in Fig. \ref{p_vs_L}, the intensity of the secondary peak decreases during the rising phase, remains constant at a low value and then increases during the falling phase as the flux gradually increases. If we consider the luminosity at which the state transition in HID occurs during the rising phase as the critical luminosity, $L \sim 1.1\times10^{37}\,$erg\,$\rm{s}^{-1}$, we observe that the pulse profile shape transition also occurs at this luminosity. In several giant outbursts of accreting pulsars, it has been noted that the transition of the pulse shapes  could serve as a hint of critical luminosities; for instance, 2S 1417-624 \citep{2020MNRAS.491.1851J}, 1A 0535+262 \citep{2022ApJ...935..125W}, and Swift~J0243.6+6124 \citep{2018MNRAS.479L.134T}. This is because the observed pulse profile shape is largely determined by the configuration of the emission region, which depends on the accretion rate. The process of pulse profile formation is complicated and hard to describe using only simple models. However, it is natural to expect a pencil beam pattern from accretion with hot spots on the neutron star surface at sub-critical mass accretion rates. At super-critical mass accretion rates, collisionless shocks and radiation-dominated accretion columns would produce a fan beam pattern \citep{2023hxga.book..138M}. If the pulse profile shape transitions are indeed due to \src\ reaching a critical luminosity, we can infer that the geometries of the emitting region has always been a mixture of pencil beam and fan beam. The main peak is dominated by the fan beam, while the secondary peak is dominated by the pencil beam, with the pencil beam contributing more at lower luminosities compared to higher luminosities. 

For typical neutron star parameters, \citet{2012A&A...544A.123B} derived the critical luminosity as $L_{\rm crit}$ $\sim$ $1.5\times 10^{37} B_{12}^{16/15}$ erg\,$\rm{s}^{-1}$, where $B_{12}$ is the surface magnetic field strength in units of $10^{12}$\,G. The value of  $L_{\rm crit}$ $\sim$ 1.1 $\times$ $10^{37}\,$erg\,$\rm{s}^{-1}$ corresponds to a magnetic field of 0.7 $\times$ $10^{12}$\,G. \citet{2015MNRAS.447.1847M} provided a numerical solution by calculating the luminosity for two scenarios: one with purely extraordinary polarization and another with an equal mix of ordinary and extraordinary polarization. The $L_{\rm crit}$ value can only be achieved in the purely extraordinary polarization case. Since the real critical luminosity value  is likely to lie between the two cases (as shown in  Fig. 7 of that study), the expected cyclotron energy ($E_{\rm cycl}$) would be approximately smaller than 5 keV and larger than 30 keV. This corresponds to a magnetic field strength lower than 0.6 $\times$$10^{12}$\,G and higher than 3.4 $\times$$10^{12}$\,G. Using \textit{NICER} data, \citet{2024A&A...688A.213T} used relativistic raytracing to model pulse profiles, showing how complex accretion column geometries affect changes in pulse profiles across a range of luminosities. They suggested a critical bolometric luminosity of 2.3 $\times$ $10^{36}\,$erg\,$\rm{s}^{-1}$ for the spectral transition and 7.9 $\times$ $10^{36}\,$erg\,$\rm{s}^{-1}$ for the transition seen in the pulse profiles, deriving a surface magnetic field of 0.17–0.55 $\times$$10^{12}$\,G. These values are slightly lower than the critical luminosity determined similarly from the \hxmt\ data. This is probably due to the different energy range used.

The pulse profile also evolved with energy. Observations of the source close to the maximum of the outburst allowed us to perform a detailed study of this dependence (Fig. \ref{PP_E}). The low- and high-energy profiles have a very complex structure of up to four peaks at lower energies that eventually merges to become a one-peaked profile at higher energies. The strong energy dependence of the pulse profiles has also been observed from the 2006 outburst by \citep{2008A&A...491..833K}, using INTEGRAL observations in the 2$-$120~keV energy range. In general, it is a common property for accreting pulsars that they show pulse profiles with a more complex shape at lower energies, while a simple one-peak or two-peak profile at higher energies (see, e.g., \citealt{2013ApJ...763...79M,2020MNRAS.491.1851J,2022ApJ...935..125W}). Variations of pulse profiles with the energy can be caused by changes in the beam pattern on photon energy and/or local absorption due to the non-spherically symmetric distribution of matter within the system. The presence of one-peaked pulses at high energies can be attributed to a fan beam pattern in the emission geometry \citep{2022ApJ...938..149H}. Conversely, interpreting complex pulse profiles at low energies is more challenging due to their susceptibility to scattering and absorption within the neutron star's local environment. 

{The energy dependence of the PF has been extensively studied in a large sample of transient Be/X-ray binary pulsars by \citet{2009AstL...35..433L}, including \src. \citet{2009AstL...35..433L} observed that \src\ exhibited a consistent increase in PF with energy during its low state, a behavior similarly reported for several other pulsars. Additionally, \citet{2008A&A...491..833K} observed a bump-like structure around 10--20~keV at peak luminosity that is probably related to the continuum feature they found in this energy range. In our study, we observed a similar bump around 10~keV near peak luminosity, which was not present in lower luminosity states. This suggests that the bump may be luminosity-dependent, warranting further observations of \src\ for confirmation.} Additionally, the PF shows an anti-correlation with luminosity during both the rise and decay phases (see Fig. \ref{PF_L}). 

Prior spectral studies of \src\ show some complex absorption features in its spectrum and cannot be modeled using a simple cutoff power-law model. \citet{1999MNRAS.302..700R} detected an additional blackbody component with $kT_{\rm bb}\,$ = 1--1.4~keV. and a possible CRSF at $\sim$36~keV. The radius of the X-ray emission surface is found to be $\sim$1\,km. By using the updated distance of 2.4\,kpc, the peak luminosity of their observation is of the order of $\sim$$10^{36}\,$erg\,$\rm{s}^{-1}$. The best-fit solution of the spectrum of the 2006 giant outburst (L $\sim$$10^{37}\,$erg\,$\rm{s}^{-1}$, d=2.4\,kpc) obtained by \citet{2007A&A...464L..45K} required an additional broad emission "bump" at 15~keV or the addition of two Gaussian absorption lines at $\sim$10 and $\sim$20~keV. \citet{2013A&A...551A...1R} observed a spectral transition from a negative to positive correlation in the L – $\Gamma$ diagram at 4 $\times$ $10^{36}\,$erg\,$\rm{s}^{-1}$, suggesting a potential indication of the critical luminosity. {\citet{2024A&A...688A.214B} confirms the presence of a narrow absorption feature around 10~keV, its interpretation as CRSF remains uncertain due to the absence of harmonics. They also observed a spectral hardening at lower luminosities during the 2021 outburst.} In our study, the broad band spectrum can be modeled by an absorbed cutoff power law associated with a blackbody component and an iron line at $\sim$6.6~keV. The photon index exhibited no clear evolution with luminosity, contrasting with the results from the 2006 outburst. Typically, spectral transitions are expected to be observed at critical luminosities in X-ray pulsars, reflecting changes between sub-critical and super-critical regimes \citealt{2013A&A...551A...1R, 2020ApJ...902...18K}). One possible explanation is that the large error bars at lower luminosities make the spectral variability less evident. Additionally, the luminosity observed during the 2006 outburst by \citet{2013A&A...551A...1R} ranges from $2 \times 10^{35}$ erg s$^{-1}$ to $2 \times 10^{37}$ erg s$^{-1}$, which is broader than that observed during the 2021 outburst by \hxmt. It might also contribute to the absence of the spectral transition.

We see the strong energy dependence of the pulse profile shape in Fig. \ref{prs}. The spectral variation with pulse phase is a common feature in X-ray pulsars (see e.g. \citealt{2008A&A...482..907K, 2015MNRAS.448.2175L}). Phase-resolved spectroscopy of \src\ was also performed in earlier works. \citet{2008A&A...491..833K} detected a Gaussian absorption line at $\sim$ 63.6~keV between the main and secondary peaks, corresponding to a phase bin of 0.4$-$0.5 in our study. This could indicate the presence of a CRSF, which are sometimes detected at higher significance in correspondence of some specific pulse phases. However, no similar feature was detected in the 2021 outburst. During the 2006 giant outburst, a minimum value of the power-law photon index was observed around the main peak. The cutoff energy exhibited a slight shift with respect to the peak towards earlier pulse phases. In the 2021 outburst, the maximum of the power law photon index and the cutoff energy coincide with the main pulse (Fig.\ref{prs}). {During the 2022 outburst, the source was observed by \ixpe, \hxmt, and ART-XC across the 2$-$70~keV energy range, at a distance of 2.4~kpc. The observed source luminosity was $\sim3\times10^{36}$\,erg\,s$^{-1}$ (\citet{2023A&A...675A..29M}, Table 2). Additionally, they reported a maximum value of the photon index around the same peak in the phase-resolved results (\citet{2023A&A...675A..29M}, Fig. 7). We emphasize that the peak luminosity of the 2006 outburst was reported as $\sim1.1\times10^{38}$\,erg\,s$^{-1}$ assuming d=7.1\,kpc (corresponding to $\sim1.3\times10^{37}$\,erg\,s$^{-1}$ if d=2.4\,kpc) from \citet{2008A&A...491..833K}, as observed with INTEGRAL JEM-X in the 2$-$9~keV energy range.} In contrast, our peak luminosity for the 2021 outburst is $\sim1.45\times10^{37}$\,erg\,s$^{-1}$, measured across the 0.01$-$150~keV range. The 2006 outburst's luminosity would likely appear greater if observed across a broader energy range. As observed by \emph{Swift}/BAT\footnote{\url{https://integral.esac.esa.int/bexrbmonitor/Plots/sim_plot_EXO2030+375.html}}, the flux of the 2006 outburst was 1.2~Crab, while the 2021 outburst flux was 0.6~Crab, indicating it was half as bright as the 2006 outburst. The contrasting phase-resolved results could be attributed to differences in luminosities.

\section{Conclusions} \label{section_5}

We  performed a comprehensive timing and spectral study of Be/X-ray binary \src. The pulse profiles exhibit strong energy and luminosity dependence, implying that changes are taking place in the geometry of emitting region. The hardness-intensity diagrams (7$-$10~keV/4$-$7~keV) indicate  transitions in state during the early and late phases of the outburst. These state transitions were aligned with the luminosity levels, where changes in the pulse profile shape occurred, indicating a critical luminosity of 1.1 $\times$ $10^{37}\,$erg\,$\rm{s}^{-1}$; however, the spectral properties did not exhibit a clear transition throughout the outburst. Additionally, no clear evidence of CRSFs was found in the phase-averaged and phase-resolved spectra during the outburst. {Future detection of CRSFs in \src\ could significantly enhance our understanding of the pulsar's magnetic field structure and its evolution during outbursts, as well as provide crucial insights for determining its critical luminosity.}

\begin{acknowledgements}
 This work made use of data from the \hxmt\ mission, a project funded by China National Space Administration (CNSA) and the Chinese Academy of Sciences (CAS). Y.J. Du would like to thank the support from China Scholarship Council (CSC 202108080247). P. J. Wang is grateful for the financial support provided by the Sino-German (CSC-DAAD) Postdoc Scholarship Program (57678375). L. D. Kong is grateful for the financial support provided by the Sino-German (CSC-DAAD) Postdoc Scholarship Program (57607866). L. Ji is supported by the National Natural Science Foundation of China under grant No. 12173103. LD acknowledges funding from the Deutsche Forschungsgemeinschaft (DFG, German Research Foundation) - Projektnummer 549824807.
\end{acknowledgements}



\begin{thebibliography}{}
  \bibitem[Arnaud(1996)]{1996ASPC..101...17A} Arnaud, K.~A.\ 1996, Astronomical Data Analysis Software and Systems V, 101, 17
  \bibitem[Bailer-Jones et al.(2021)]{2021AJ....161..147B} Bailer-Jones, C.~A.~L., Rybizki, J., Fouesneau, M., et al.\ 2021, \aj, 161, 147
  \bibitem[Ballhausen et al.(2024)]{2024A&A...688A.214B} Ballhausen, R., Thalhammer, P., Pradhan, P., et al.\ 2024, \aap, 688, A214
  \bibitem[Basko \& Sunyaev(1976)]{1976MNRAS.175..395B} Basko, M.~M. \& Sunyaev, R.~A.\ 1976, \mnras, 175, 395
  \bibitem[Becker et al.(2012)]{2012A&A...544A.123B} Becker, P.~A., Klochkov, D., Sch{\"o}nherr, G., et al.\ 2012, \aap, 544, A123
  \bibitem[Cao et al.(2020)]{2020SCPMA..6349504C} Cao, X., Jiang, W., Meng, B., et al.\ 2020, Science China Physics, Mechanics, and Astronomy, 63, 249504
  \bibitem[Chen et al.(2020)]{2020SCPMA..6349505C} Chen, Y., Cui, W., Li, W., et al.\ 2020, Science China Physics, Mechanics, and Astronomy, 63, 249505
  \bibitem[Ferrigno et al.(2016)]{2016xnnd.confE..44F} Ferrigno, C., Pjanka, P., Bozzo, E., et al.\ 2016, XMM-Newton: The Next Decade, 44
  \bibitem[Fu et al.(2023)]{2023MNRAS.521..893F} Fu, Y.-C., Song, L.~M., Ding, G.~Q., et al.\ 2023, \mnras, 521, 893
  \bibitem[Hou et al.(2022)]{2022ApJ...938..149H} Hou, X., Ge, M.~Y., Ji, L., et al.\ 2022, \apj, 938, 149
  \bibitem[Ji et al.(2019)]{2019MNRAS.484.3797J} Ji, L., Staubert, R., Ducci, L., et al.\ 2019, \mnras, 484, 3797
  \bibitem[Ji et al.(2020)]{2020MNRAS.491.1851J} Ji, L., Doroshenko, V., Santangelo, A., et al.\ 2020, \mnras, 491, 1851
  \bibitem[Klochkov et al.(2007)]{2007A&A...464L..45K} Klochkov, D., Horns, D., Santangelo, A., et al.\ 2007, \aap, 464, L45
  \bibitem[Klochkov et al.(2008)]{2008A&A...482..907K} Klochkov D., Staubert R., Postnov K., et al.\ 2008, \aap, 482, 907
  \bibitem[Klochkov et al.(2008)]{2008A&A...491..833K} Klochkov, D., Santangelo, A., Staubert, R., et al.\ 2008, \aap, 491, 833
  \bibitem[Kong et al.(2020)]{2020ApJ...902...18K} Kong, L.~D., Zhang, S., Chen, Y.~P., et al.\ 2020, \apj, 902, 18
  \bibitem[Kong et al.(2021)]{2021ApJ...917L..38K} Kong, L.~D., Zhang, S., Ji, L., et al.\ 2021, \apjl, 917, L38
  \bibitem[Leahy(1987)]{1987A&A...180..275L} Leahy, D.~A.\ 1987, \aap, 180, 275
  \bibitem[Li et al.(2020)]{2020JHEAp..27...64L} Li, X., Li, X., Tan, Y., et al.\ 2020, Journal of High Energy Astrophysics, 27, 64
  \bibitem[Liao et al.(2020)]{2020JHEAp..27...14L} Liao, J.-Y., Zhang, S., Lu, X.-F., et al.\ 2020, Journal of High Energy Astrophysics, 27, 14
  \bibitem[Liu et al.(2020)]{2020SCPMA..6349503L} Liu, C., Zhang, Y., Li, X., et al.\ 2020, Science China Physics, Mechanics, and Astronomy, 63, 249503
  \bibitem[Lutovinov \& Tsygankov(2009)]{2009AstL...35..433L} Lutovinov, A.~A. \& Tsygankov, S.~S.\ 2009, Astronomy Letters, 35, 433
  \bibitem[Lutovinov et al.(2015)]{2015MNRAS.448.2175L} Lutovinov, A.~A., Tsygankov, S.~S., Suleimanov, V.~F., et al.\ 2015, \mnras, 448, 2175
  \bibitem[Malacaria et al.(2023)]{2023A&A...675A..29M} Malacaria, C., Heyl, J., Doroshenko, V., et al.\ 2023, \aap, 675, A29
  \bibitem[Maraschi et al.(1976)]{1976Natur.259..292M} Maraschi, L., Treves, A., \& van den Heuvel, E.~P.~J.\ 1976, \nat, 259, 292
  \bibitem[Maitra \& Paul(2013)]{2013ApJ...763...79M} Maitra, C. \& Paul, B.\ 2013, \apj, 763, 79
  \bibitem[Matsuoka et al.(2009)]{2009PASJ...61..999M} Matsuoka, M., Kawasaki, K., Ueno, S., et al.\ 2009, \pasj, 61, 999
  \bibitem[Mushtukov et al.(2015)]{2015MNRAS.447.1847M} Mushtukov, A.~A., Suleimanov, V.~F., Tsygankov, S.~S., et al.\ 2015, \mnras, 447, 1847
  \bibitem[Mushtukov \& Tsygankov(2023)]{2023hxga.book..138M} Mushtukov, A. \& Tsygankov, S.\ 2023, Handbook of X-ray and Gamma-ray Astrophysics, 138. 
  \bibitem[Parmar et al.(1985)]{1985IAUC.4066....1P} Parmar, A.~N., Stella, L., Ferri, P., et al.\ 1985, \iaucirc, 4066
  \bibitem[Parmar et al.(1989)]{1989ApJ...338..359P} Parmar, A.~N., White, N.~E., Stella, L., et al.\ 1989, \apj, 338, 359
  \bibitem[Reig(2011)]{2011Ap&SS.332....1R} Reig, P.\ 2011, \apss, 332, 1
  \bibitem[Reig \& Coe(1999)]{1999MNRAS.302..700R} Reig, P. \& Coe, M.~J.\ 1999, \mnras, 302, 700
  \bibitem[Reig \& Nespoli(2013)]{2013A&A...551A...1R} Reig, P. \& Nespoli, E.\ 2013, \aap, 551, A1
  \bibitem[Staubert et al.(2014)]{2014A&A...572A.119S} Staubert, R., Klochkov, D., Wilms, J., et al.\ 2014, \aap, 572, A119
  \bibitem[Staubert et al.(2019)]{2019A&A...622A..61S} Staubert, R., Tr{\"u}mper, J., Kendziorra, E., et al.\ 2019, \aap, 622, A61
  \bibitem[Tamang et al.(2022)]{2022MNRAS.515.5407T} Tamang, R., Ghising, M., Tobrej, M., et al.\ 2022, \mnras, 515, 5407
  \bibitem[Thalhammer et al.(2024)]{2024A&A...688A.213T} Thalhammer, P., Ballhausen, R., Sokolova-Lapa, E., et al.\ 2024, \aap, 688, A213
  \bibitem[Tsygankov et al.(2017)]{2017A&A...605A..39T} Tsygankov, S.~S., Doroshenko, V., Lutovinov, A.~A., et al.\ 2017, \aap, 605, A39
  \bibitem[Tsygankov et al.(2018)]{2018MNRAS.479L.134T} Tsygankov, S.~S., Doroshenko, V., Mushtukov, A.~A., et al.\ 2018, \mnras, 479, L134
  \bibitem[Wang et al.(2022)]{2022ApJ...935..125W} Wang, P.~J., Kong, L.~D., Zhang, S., et al.\ 2022, \apj, 935, 125
  \bibitem[Wilms et al.(2000)]{2000ApJ...542..914W} Wilms, J., Allen, A., \& McCray, R.\ 2000, \apj, 542, 914
  \bibitem[Zhang et al.(2020)]{2020SCPMA..6349502Z} Zhang, S.-N., Li, T., Lu, F., et al.\ 2020, Science China Physics, Mechanics, and Astronomy, 63, 249502
\end{thebibliography}

\begin{appendix}
\section{Observations}\label{sec:observations}

\begin{table}[H]

\caption{\hxmt observations of \src\ during the 2021 outburst used for data analysis. \label{tab:HXMTobsID_rise&decay}}
\centering
\renewcommand{\arraystretch}{0.8}
\begin{adjustbox}{max width=\columnwidth}

\begin{tabular}{ccccc}
\hline\hline
Observation ID & Date (MJD) &\multicolumn{3}{c}{Exposure (s)} \\ 
\cline{3-5} 
& & {LE } & {ME } & {HE }  \\ 
\hline
03 & 59427.23 & 1725 & 2522 & 3579 \\
04 & 59429.09 & 2274 & 4306 & 4118 \\
05 & 59430.55 & 6022 & 6170 & 5537 \\
07 & 59432.46 & 4847 & 5856 & 5169 \\
08 & 59433.45 & 3950 & 6018 & 5166 \\
09 & 59434.73 & 2785 & 3958 & 4886 \\
11 & 59436.78 & 3691 & 6231 & 4137 \\
12 & 59437.78 & 5053 & 6475 & 2986 \\
13 & 59438.84 & 6357 & 6646 & 5468 \\
16 & 59441.69 & 6394 & 7558 & 5320 \\
18 & 59445.37 & 4999 & 7817 & 6148 \\
19 & 59447.52 & 2439 & 4625 & 6430 \\
21 & 59451.62 & 4842 & 7544 & 6233 \\
23 & 59453.48 & 4949 & 7944 & 6268 \\
24 & 59454.93 & 1705 & 2206 & 2825 \\
25 & 59457.92 & 958 & 2575 & 1752 \\
107\tablefootmark{a} & 59460.11 & 299 & 1696 & 492\\
108\tablefootmark{a} & 59460.24 & 946 & 3328 & 4236 \\
109\tablefootmark{a} & 59460.38 & 416 & 2800 & 1777 \\
110\tablefootmark{a} & 59460.51 & 1257 & 2384 & 1403 \\
111\tablefootmark{a} & 59460.64 & 2095 & 2165 & 2859 \\
112\tablefootmark{a} & 59460.77 & 1822 & 1963 & 2581 \\
113\tablefootmark{a} & 59460.91 & 599 & 656 & 874 \\
26 & 59461.74 & 3742 & 3440 & 3974 \\
27 & 59463.47 & 6791 & 7781 & 7543 \\
28 & 59464.85 & 838 & 1643 & 1376 \\
29 & 59465.74 & 1525 & 1521 & 2013 \\
30 & 59466.98 & 1076 & 2241 & 1865 \\
31 & 59467.84 & 2334 & 3785 & 4244 \\
32 & 59468.69 & 2035 & 2495 & 2765 \\
33 & 59469.82 & 2274 & 3406 & 3569 \\
34 & 59471.01 & 5686 & 8356 & 6145 \\
35 & 59473.13 & 5686 & 8084 & 5745 \\
36 & 59475.81 & 6118 & 12074 & 8997 \\
37 & 59477.18 & 4317 & 11427 & 10438 \\
38 & 59479.17 & 3721 & 4417 & 4346 \\
39 & 59481.16 & 6118 & 4465 & 5414 \\
40 & 59483.29 & 3967 & 4993 & 6851 \\
41 & 59486.01 & 1496 & 1786 & 1340 \\
42 & 59488.32 & 1887 & 3548 & 4766 \\
43 & 59489.31 & 4856 & 7150 & 6345 \\
44 & 59491.30 & 1886 & 3851 & 4874 \\
45 & 59493.18 & 3651 & 5752 & 6963 \\
46 & 59495.09 & 3855 & 6016 & 7551 \\
47 & 59497.08 & 1647 & 6496 & 6929 \\
48 & 59499.06 & 3015 & 7020 & 7916 \\
51 & 59506.08 & 1409 & 5484 & 5381 \\
52 & 59507.48 & 958 & 3672 & 2787 \\
54 & 59511.05 & 1197 & 7559 & 6635 \\
55 & 59513.08 & 2769 & 8764 & 6949 \\
56 & 59515.29 & 3032 & 5933 & 7083 \\
59 & 59521.15 & 7401 & 10512 & 6292 \\
60 & 59523.16 & 7336 & 11385 & 9217 \\
61 & 59525.20 & 4144 & 8897 & 6228 \\
62 & 59527.13 & 3084 & 5364 & 6263 \\
63 & 59529.05 & 8411 & 8789 & 8060 \\
64 & 59531.10 & 6026 & 7309 & 6141 \\
65 & 59533.15 & 4992 & 6941 & 5169 \\
66 & 59535.06 & 4927 & 7133 & 5558 \\
\hline
\end{tabular}
\end{adjustbox}
\tablefoot{
For each observation ID, P03040300 should be prefixed. For observation IDs marked with (a) in the upper right corner, P040414700 should be prefixed, and 20210903-02-01 should be appended.
}
\end{table}

\section{Pulse-on and pulse-off spectral analysis}\label{sec:pulse-on-off}

\begin{figure}
\centering
\includegraphics[width=9cm]{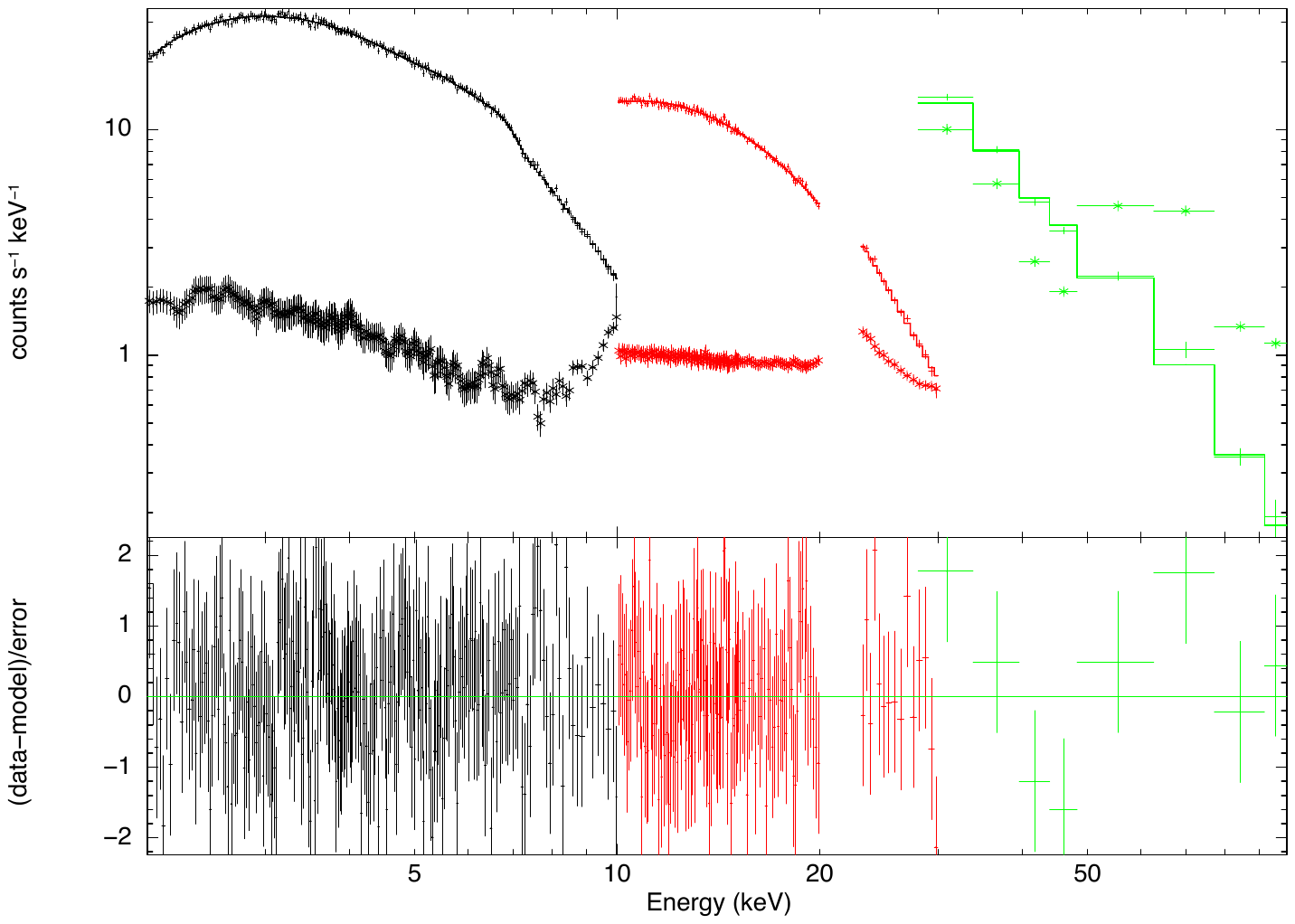}
\caption{Broadband spectrum of EXO 2030+375 (ObsID: P030403002703) in energy range 2$-$100~keV using the model \texttt{const$\times$tbabs(gauss+bbodyrad+gauss+cutoffpl)}. The asterisk marks represent background values, the solid crosses represent net spectral values, and the step line indicates the model.}
\label{1--100}
\end{figure}

\begin{figure}
\centering
\includegraphics[width=9cm]{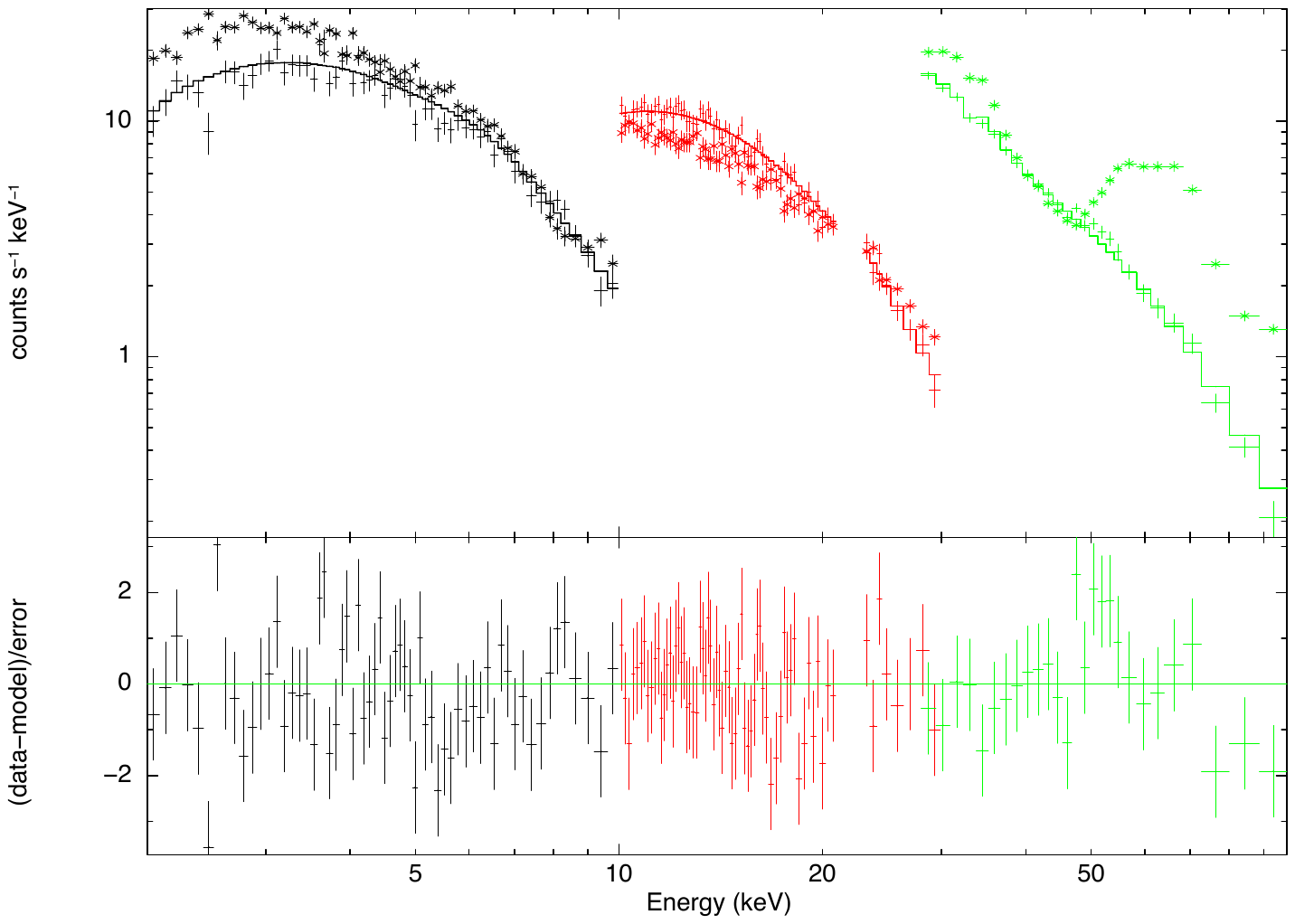}
\caption{Pulse-on minus pulse-off spectrum of ObsID P03040300270 in the energy range 2--100~keV using the model \texttt{const$\times$tbabs$\times$cutoffpl}.}
\label{pulseon/off}
\end{figure}

We wish to highlight our decision to restrict the energy range to 2$-$50 keV for the spectral analysis, aiming to focus on an energy range with reliable background calibration. Initially, we analyzed data ranging from 2$-$100~keV. {As illustrated in Fig \ref{1--100}, the asterisk marks denote the background values for ObsID P030403002703. The characteristic "U-shaped" background (see in \citep{2020JHEAp..27...14L} for the description of the background model) observed between 40~keV and~50 keV, may be responsible for the feature seen around 47~keV.} For a more thorough investigation of the feature, we performed pulse-on and pulse-off spectrum analysis for ObsID P030403002703. {This analysis involves comparing the X-ray spectra at different phases of the pulsar's rotation, specifically when the pulsar's emission is strongest (pulse-on) and weakest (pulse-off). By using the spectrum from the weakest phase as the background, this method helps avoid the influence of an excessively high background, which can otherwise obscure the true emission features. }The phase bin extracted for the pulse-on spectrum covered phases 0.5-0.7 (mostly encompassing the main peak), while the phase bin for the pulse-off spectrum covered phases 0.97-1.03 (centered on the lowest flux). As seen in Fig \ref{pulseon/off}, we did not observe the 47~keV feature and the background is significant above 50~keV. Therefore, we restricted the energy range to 2$-$50~keV in this paper to minimize uncertainties associated with background. 

\end{appendix}
\end{document}